\documentclass[12pt]{iopart} 
\input psfig.sty
 
\begin{document} 
 
\title[Mechanisms and sites of ultra high energy cosmic ray 
origin]{Mechanisms and sites of ultra high energy cosmic ray
origin}
 
\author{M. Ostrowski} 
\address{Obserwatorium Astronomiczne, Uniwersytet Jagiello\'nski, \\
ul. Orla 171, 30-244 Krak\'ow, Poland (e-mail: mio\@@oa.uj.edu.pl)}
 
%\maketitle 
 
\begin{abstract} 
 
We shortly discuss several astrophysical scenarios leading to cosmic ray
acceleration up to extremely high energies reaching the scale of
$10^{20}$ eV. The processes suggested in the literature include
acceleration at relativistic jet terminal shocks and shear boundary
layers, shocks in large scale accretion flows onto supergalactic cosmic
structures, particle reflections from ultra-relativistic shocks
postulated to exist in sources of gamma ray bursts, the processes
involving the neutron star rotating magnetospheres and dormant quasars.
Some of these objects can explain cosmic rays with highest energies if
one tunes the model parameters to limits enabling the highest
acceleration efficiency. We also note that some of the considered
processes allow for acceleration efficiency in the Hillas diagram,
$\beta$, to be much larger than unity. The present paper is based on a
review talk presented during the European Cosmic Ray Symposium in Lodz
(2000).
\end{abstract}

\section{Introduction} 
 
Present search for explanation of the observed ultra high energy 
(UHE) cosmic rays (CRs) resembles somewhat a random walk among possible 
and impossible particle acceleration models and wide range of more or 
less exotic physical concepts and particles. The expanding space of such 
studies is described by more and more numerous review papers appearing 
in ASTRO-PH listings. The present one is limited to discussing 
some astrophysical mechanisms proposed to accelerate cosmic rays up to the
highest observed energies. In our opinion necessity to seriously 
consider sources involving `strange' particles or new physics will come 
after astrophysical models applying the classical physics are rejected by
observations. A review of UHE CR observations is presented by 
e.g. Bertou et al. [5]. 
 
The factors limiting energies of particles accelerated in astrophysical 
sources include an acceleration rate and a loss rate due to 
synchrotron radiation and IC or inelastic scattering. However, for the 
highest considered energies quite often an upper energy limit is 
provided by a finite source extent. To evaluate the role of this last 
factor one needs information about magnetic fields present within its 
volumes. For objects with relatively strong magnetic fields such
data are quite often firmly constrained by radio and high energy
observations. Contrary to that, knowledge of weak magnetic fields in
extragalactic space or in galactic coronae, important for the discussed
below models involving supergalactic accretion shocks or local sources
concentrated toward the galaxy centre, is only fragmentary. Therefore,
let us start with a short review of the new evidence on such magnetic
fields, which could also substantially influence transport of $\sim
10^{20}$ eV particles reaching Earth. Next, we will discuss a few
astrophysical `bottom-up' models for generation of such particles. {\it Our
intention is not to discuss the full literature of the subject, but
rather to overview a set of characteristic publications presenting the
discussed models.}

Below, in most cases we use a term `particle' for protons.

\section{Extragalactic magnetic fields} 
 
The importance of any considered magnetized structure for particle
acceleration can be evaluated by comparing its size to the particle
gyroradius\footnote{We often use the notation of the form $X_Y$ for a
quantity $X$ divided by $10^Y$ of the respective units, e.g. for energy
$E_{18} \equiv E / 10^{18}$ eV, for the magnetic field $B_{-6} \equiv B/
10^{-6}$ G, or for a distance $R_6 \equiv R/10^6$ cm.}

$$ r_g \approx { 1 \over Z } {E_{18} \over B_{-6} } \quad {\rm kpc}
\qquad , \eqno(1)$$
 
\noindent 
where $Z$ is a given nucleus charge state. Early evaluations of the 
field strength at {\it largest scales} yielded usually upper limits only of 
the magnitude order $10^{-9}$~G. The direct observations of extended 
extragalactic synchrotron glows started in 1990, and were continued in 
last few years, revealing much stronger fields within galaxy clusters 
and even in the intracluster space (for review see Kronberg [13]). 
Estimates show, that $B$ may reach $\sim 1$ $\mu$G in rich clusters and 
approximately $\sim 0.1$ $\mu$G in supergalactic accretion flows and in 
supergalactic structures at several Mpc scales. Turning to smaller scales 
a series of 
spiral galaxy observations proves existence of extended  -- with the
vertical scale $\sim$ several kpc -- magnetic coronae near some (not 
all) galaxies with a few $\mu$G magnetic fields (cf. Beck [2]). 
Even irregular galaxies show well organized magnetic field structures 
(Chy\.zy et al. [8]). Growing evidence supporting the Kronberg's claim, 
that magnetic fields appear to be all-pervasive in space, weaken the 
opinion that the UHE cosmic ray showers' directions point approximately 
to sources of the observed particles (cf. Sigl et al. [18]). However, 
there exists also a growing number of observed particle pairs and 
triplets above a few times $10^{19}$ eV which could suggest the opposite 
(cf. Uchihori et al. [20]).
 
\section{Particle acceleration in relativistic jets} 
 
Relativistic jets occurring in FRII radio galaxies carry large energy up 
to the radio `hot spots' situated far ($\sim 100$ kpc) from the central 
source. These hot spots are believed to harbor strong, mildly 
relativistic shocks dissipating the jet bulk kinetic energy into heating 
plasma, generating magnetic fields and efficiently accelerating 
energetic particles. Because of relatively slow radiative losses for 
protons such shocks are prospectus UHE cosmic ray accelerators. 
Additionally,  an active role in accelerating UHE cosmic rays can play a 
velocity shear layer at the relativistic jet side boundary. Let us discuss 
these possibilities in some detail. 
 
\subsection{Relativistic shocks in hot spots of FRII radio galaxies} 
 
A full elaborated model of UHE CRs acceleration at the relativistic jet
terminal shock was presented by, e.g., Rachen \& Biermann [17]. They
considered a shock wave with parameters derived from observations of 
hot spots in the considered sources. One can model the downstream region
of the shock to be a circular slice of compressed plasma with the radius 
in the range $R \approx 0.3 \, - \, 3$ kpc, thickness $H \approx 0.1 \, 
- \, 2$ kpc, the magnetic field $B \approx 0.4 \, - \, 0.6$
mG. The evaluated jet velocities are in the mildly-relativistic range
$u_j \approx 0.2 \, - \, 0.5$ c. In order to derive the upper limit for 
the accelerated proton energy, $E_{max}$,  we compare the shock 
acceleration time scale given by Rachen \& Biermann for a parallel 
`$\|$' shock 
 
$$\tau_{acc} = {20 \kappa_\| \over u_j^2 } \qquad , \eqno(2)$$ 
 
\noindent 
where $\kappa_\|$ is the cosmic ray diffusion coefficient parallel to the 
mean magnetic field, with the loss time scale scaled to that of the
synchrotron radiation
 
$$\tau_{loss} = { C \over B^2 (1+X) \gamma_p } \qquad , \eqno(3)$$ 
 
\noindent 
where $C$ ($\approx 5 \cdot 10^{24}$s for $B$ given in mG) is a constant,
$X$ represents (in some conditions a large, varying with the local
conditions and particle energy) correction to $\tau_{loss}$ due to inverse-Compton (`IC') 
and inelastic collisions, and $\gamma_p$ is a proton Lorentz factor.
Near the shock, Rachen \& Biermann [17] considered the non-linear
Kolmogorov turbulence extended up to the scales comparable to the hot
spot size. Thus, in the long wave range important for the highest energy
particle scattering the diffusion was the Bohm diffusion, $\kappa_\|
\approx {1 \over 3} r_g c$, leading to the most rapid acceleration. Then,
for `typical' hot-spot parameters $B = 0.5$ mG, $u_j = 0.3$ c, $R > H
\approx 1$ kpc, $X < 1$, the conditions $\tau_{acc} < \tau_{loss}$ and
$r_g < H$ can be satisfied up to energies of a few $10^{20}$ eV.
 
However, some of the above assumptions or evaluations are only rough 
estimates, which make the derived $E_{max}$ somewhat uncertain. For 
example a required diffusive size for $\sim 10^{20}$ eV particles
seems to be at least (and in fact more than) $(c/u_j) r_g$ (cf. Drury
[9]), otherwise the particle spectrum exhibits a cut-off (see also
below). Also, the turbulence structure downstream of the shock can
substantially deviate from the assumed Kolmogorov form.

\subsection{Acceleration at the jet shear boundary layer} 
 
Ostrowski ([14, 15]) discussed the process of particle acceleration up 
to ultra high energies at tangential velocity transitions at side boundaries 
of relativistic jets. An UHE particle can cross such boundary to inside or 
to outside the jet, then be scattered back to cross the jet boundary 
again. If the process repeats, each boundary crossing increases particle 
energy by, on average,
 
$${\Delta E \over E} = \eta_E \, (\Gamma - 1) \qquad , \eqno(4)$$ 
 
\noindent 
where $\Gamma$ is the jet Lorentz factor and the efficiency factor 
$\eta_E$ depends on particle anisotropy at the boundary and, thus, on the 
character of MHD turbulence responsible for particle scattering. 
For highly turbulent 
conditions $\eta_E$ can be a substantial fraction of unity, decreasing 
like $\sim \Gamma^{-1}$ for larger $\Gamma$. As a result high energy 
particles near the jet boundary can be accelerated forming a power-law 
spectrum up to some cut-off energy, $E_c$, appearing when the respective 
particle gyroradius $r_g(E_c)$ becomes comparable to the jet radius, 
$R_j$. The performed simulations show that $E_{max}$ obtained in the 
boundary acceleration process is comparable to the discussed above 
maximum particle energy obtained at the terminal shock. It grows with
increasing $u_j$, but always $r_g(E_c) < R_j$: in the simulations for 
mildly relativistic jets usually $r_g(E_c) < 0.1 R_j$. An additional 
interesting feature of this acceleration process is the fact, that 
particles {\it escaping} diffusively from the jet vicinity can posses an 
extremely hard spectrum, with the power concentrated in particles near 
$E_c$. 
 
The analogous acceleration processes can act at ultrarelativistic jets 
in blazars (Ostrowski [15]). However, due to expected decrease of 
acceleration efficiency with particle anisotropy, $\eta_E \propto 
\gamma^{-1}$, the highest energies are still limited by the geometric 
condition $r_g(E_c) < R_j$. For characteristic conditions derived
for such jets, $B \sim 1$~G and $R_j \sim 10^{16}$~cm, particle energies 
above $10^{19}$~eV can be obtained. When considered as the source of UHE 
CRs escaping into the intergalactic space such processes can be further 
degraded by particle interactions with the strong ambient photon field
and/or the diffuse medium near the active galactic nucleus in the parent
galaxy. Till now these possibilities were not discussed in detail.
 
\section{Compressive flows onto supergalactic structures} 
 
Hydrodynamic modelling of cosmological structure formation yields 
extended flat or cylindrical supergalactic structures of compressed 
matter, including galaxy clusters as its sub-components. 
Diffuse plasma accreted with velocities $u \sim 10^3$~km/s at 
such extended at several Mpc structures can form large scale
shocks seen in radio observations mentioned in section 2. Considering 
UHE particle energization at such shocks (e.g. Kang et al. [12]) in the 
process of Fermi diffusive acceleration meets however serious obstacles. 
In the considered acceleration region with $\sim 0.1 \, \mu$G 
magnetic field the $10^{20}$ eV proton posses a gyroradius $r_g \sim 
1$~Mpc. Thus (cf. Drury [9]), the particle mean free path $\lambda > 
r_g$ leads to the unreasonably large diffusive region required for 
acceleration, with the size $L_{diff} > {c \over u} \lambda \sim 
300$~Mpc (!), and the acceleration time comparable to the age of 
the universe. As the considered mean free path is the one normal to the 
shock surface, $\lambda_n$, the above authors considered the so called 
`Jokipii diffusion' regime at perpendicular shocks (i.e. where the shock 
normal is perpendicular to the mean magnetic field), with $\lambda_n \ll 
r_g$. Such models formally allow for larger energies of accelerated 
particles, but can increase the upper energy limit 
in comparison to parallel shocks at most by a factor of a few (if at 
all, cf. [16] ). It is due to the fact that the Jokipii diffusion requires a medium 
amplitude perturbations of the magnetic field, and thus large mean free 
paths, $\lambda_\| \gg r_g$, or drifts along the considered shock structure. 
 
In some way analogous (super)galactic scale shock can be formed and 
accelerate particles in the observed cases of colliding galaxies 
(Al-Dargazelli et al. [1]). However, in analogy to the discussed above 
constraints the evaluation of the upper particle energy limit presented 
by these authors seems to be over optimistic. 
 
\section{Rotating neutron star magnetospheres} 
 
Extremely powerful cosmic engines are provided by rapidly rotating 
neutron stars. Of particular interest are the objects with magnetar-like 
magnetic fields ($B \gg 10^{12}$~G) and rotation periods, $T$, in the 
mili-second range. 
 
\subsection{Strong magnetic field region} 
 
A rotating magnetosphere generates an electric field of the magnitude 
$\sim {1 \over c} | ({\bf \Omega} {\rm x} {\bf R})  {\rm x} {\bf
B}|$. Multiplied by the stellar radius $R$ this field translates into a 
voltage $\Delta \Phi \sim 10^{20} \, {\rm V} \, B_{13} \, R_6^2 / T_{-3}$, which could be used for acceleration of charged nuclei. 
However, the ($e^+$, $e^-$) pairs created in the pulsar magnetosphere 
will short circuit the above potential drop to values a few orders of 
magnitude lower (cf. Venkatesan et al. [22]). Additionally, a more
realistic geometry would introduce a further decrease of $\Delta \Phi$ 
by a factor $\sim R \Omega / c \sim 0.1$. If one takes into account a possible catastrophic energy changes in the strong magnetic field and 
the radiation field near the pulsar, the co-rotating field region of the 
pulsar magnetosphere does not seem to be a possible site for extremely 
energetic nuclei acceleration (only iron or some other heavy nuclei, 
probably the not fully ionized ones, can be available there). Venkatesan
et al. [22] expect that the far field region, near the outer gap or
even behind the light cylinder, can be a more promising acceleration 
site for UHE particles. However, no detailed model of such process 
was presented till now. 
 
\subsection{Ultrarelativistic wind zone and its terminal shock} 
 
As discussed by numerous authors the rapidly rotating magnetized neutron
stars can be sources of ultra-relativistic winds, as (probably) seen in
the centre of the Crab nebula. If iron nuclei can be injected from the
star surface into the magnetosphere, then they start to co-move with the
forming wind, reaching its terminal velocity. As discussed by Blasi et
al. [6] for newly born neutron stars with $T < 10$ ms and $B \sim
10^{12} \, - \, 10^{14}$~G the wind terminal Lorentz factor can reach
$\Gamma_{wind} \sim 10^9$. Thus the iron nuclei carried with the wind
have in the `laboratory' frame typical energies $\approx 4 \cdot 10^{20}
B_{13} \Omega_3^2$~eV. At the wind terminal shock the nuclei are
injected into the interstellar medium with these energies. For sources
distributed in our Galaxy very small source efficiency is sufficient to
explain particle flux observed at Earth. Of course, if such processes
provide UHE CRs, an anisotropy toward the Galactic centre should be
observed. However, as claim the above authors, if our Galaxy posses an
extended magnetized corona the anisotropy at the northern hemisphere
pointing outside the Galactic Centre could be quite small at
$10^{20}$~eV (?). One may note that the presented model contradicts
interpretation of the Fly Eye result suggesting transition to lighter
cosmic ray composition at highest energies. Also efficiency of iron ions
pull out from the neutron star surface is a theoretically undetermined
quantity yet.
 
\section{`Dead quasars'} 
 
Observations show a much higher density of quasars in the early 
universe, at cosmological redshifts $z \approx 2$, than today. 
Thus a number of present day 
galaxies must harbour supermassive black holes in their centres,
some of them rapidly rotating. Only an extremely low accretion rate due 
to lack of freshly acquired gas near such black holes makes them to be 
inactive `dead quasars', without any clear radiative or jet 
observational signature. Nevertheless, even a small accretion rate can 
bring magnetized plasma toward a rotating black hole and to form a 
stressed magnetosphere near to it. Boldt \& Ghosh [7] estimate the 
magnetic field near the black hole horizon of a radius $R$, to reach $B 
\sim 10^4$~G and to generate an effective electromotive force $\sim cBR 
\approx 4.4 \cdot 10^{20} B_4 M_9$~V accelerating nuclei (a black hole 
mass $M_9 \equiv M/10^9$M$_o$). This force could accelerate the observed 
UHE cosmic rays in nuclei of nearby inactive massive galaxies. However, 
an ability of the small amount of accreted plasma to keep the black hole
magnetic field at $\sim 10^4$~G is not proved and should be discussed 
quantitatively. 
 
\section{Ultrarelativistic shock waves in GRBs} 
 
The observed approximately once a day gamma ray bursts are believed to 
originate from ultra-relativistic shocks, with the Lorentz $\Gamma$ 
factors reaching values $\sim 10^3$. Basing on an oversimplified
acceleration model Vietri [21] and Waxman [23] suggested that such
shocks could provide UHE CRs also. Later discussions by Gallant \&
Achterberg [10] and Bednarz \& Ostrowski [3] show that due to extreme
particle anisotropy (upstream of the shock the energetic particle
distribution has an opening angle $\sim \Gamma^{-1}$) the acceleration
process is gradual, with $\Delta E / E \sim 1$, and the considered
scenarios for GRB shock acceleration lead to particle energies up to
$\sim 10^{18}$~eV. The situation could be more promising if the shock
propagates in the region of strong magnetic field, like in the pulsar
wind zone. However no one attempted to model such acceleration process
including into consideration the pulsar wind velocity field with a
possible large Lorentz factor and the sector-like magnetic field
structure. The process of particle reflection from the shock leading in
principle to fast acceleration with $\Delta E / E \sim \Gamma^2$ is much
less efficient than thought previously (Bednarz \& Ostrowski [4]). Also
Stecker [19] pointed out that GRBs, possibly following the star
formation rate, are much less frequent in the local universe than in the
young galaxies at redshifts above 1. He evaluates that particles from
such local {\it unatenuated} sources can explain only a small part of
the observed particle flux at highest energies. Thus, in our opinion, it
is highly questionable if GRBs could generate the observed highest
energy cosmic rays.
 
\section{Concluding remarks}

\begin{figure}
\vspace{19cm}
\includegraphics{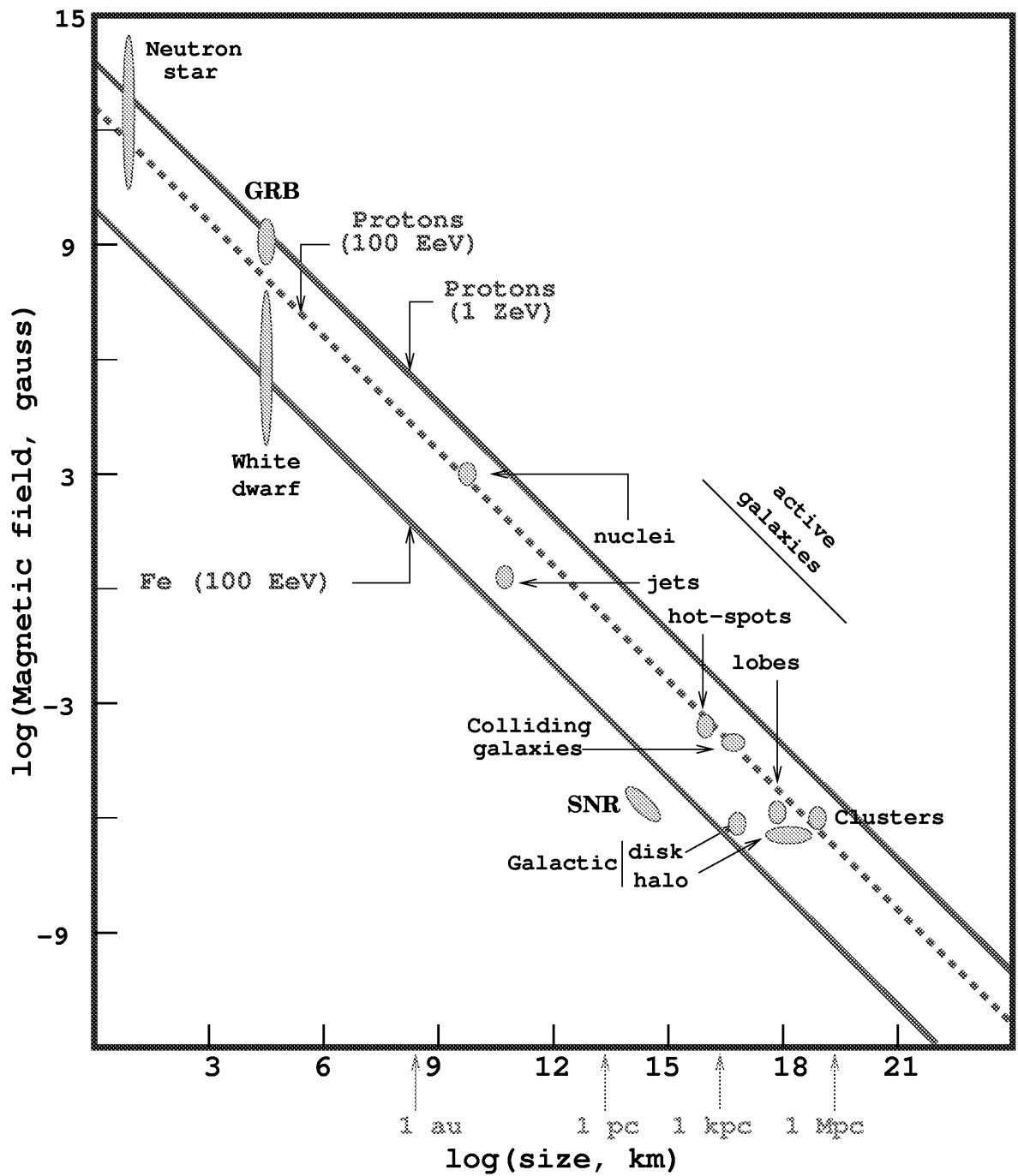}
\caption{The Hillas diagram. Acceleration of cosmic rays up to a given
energy requires conditions above the respective line. Placement of an oval 
for GRB sources is somewhat controversial at such diagram, as discussed in 
the Appendix. The figure was kindly provided by Murat Boratav. }
\end{figure}

The Hillas [11] diagram presents possible astrophysical sources of ultra
high energy particles on the plane `$B$ versus $L$' ($L$ - linear size 
of the source), where the regions of $r_g < \beta L$ roughly define the
allowed range for sources with the `acceleration parameter'
$\beta$ (Fig.~1). Hillas introduced $\beta$ as the respective velocity
parameter for the studied acceleration mechanisms. Traditional
objects placed on the diagram include galaxy clusters, AGNs, FRII radio
sources and pulsars. Recent particle acceleration models make the Hillas
diagram more densely populated due to taking into consideration objects
like `magnetars', `dead quasars', or ultrarelativistic pulsar winds and
GRB shocks. They also require a more general meaning for
the parameter $\beta$. Traditional approach considers $\beta$ to be
comparable or smaller than unity. For example particle acceleration at
relativistic shock can be characterized with $\beta \approx 1$, while
the non-relativistic shock propagating with velocity $u$ yields at 
most $\beta
\sim u/c$. In the last expression the factor $u/c$ comes due to the fact
that the source size must be greater than the required diffusive length
scale $\sim r_g c/u$ for highly turbulent conditions (cf. [9]). The
processes like a particle reflection from an ultra-relativistic shock or
a bulk flow acceleration on the cost of the pulsar wind Poynting flux
introduce a `new dimension' into the above efficiency considerations.
The model of Blasi et al. [6] considers particles resting with respect
to the pulsar wind and introduces high energy particles when injected at
the wind terminal shock. The model works well even with zero energy
particles in the wind rest frame and - in this frame - the condition
$r_g < \beta L$ does not provide any limit to particle energy in the
`external observer' rest frame. In the model involving acceleration by
cosmic ray reflections from ultra-relativistic shocks of GRBs the
condition $r_g \sim L$ downstream the shock (within the source) yields
$r_g \sim \Gamma L$ after the reflection, and thus $\beta \sim \Gamma
\gg 1$ (cf. Appendix).

\begin{table}
\vspace{14cm}
\includegraphics{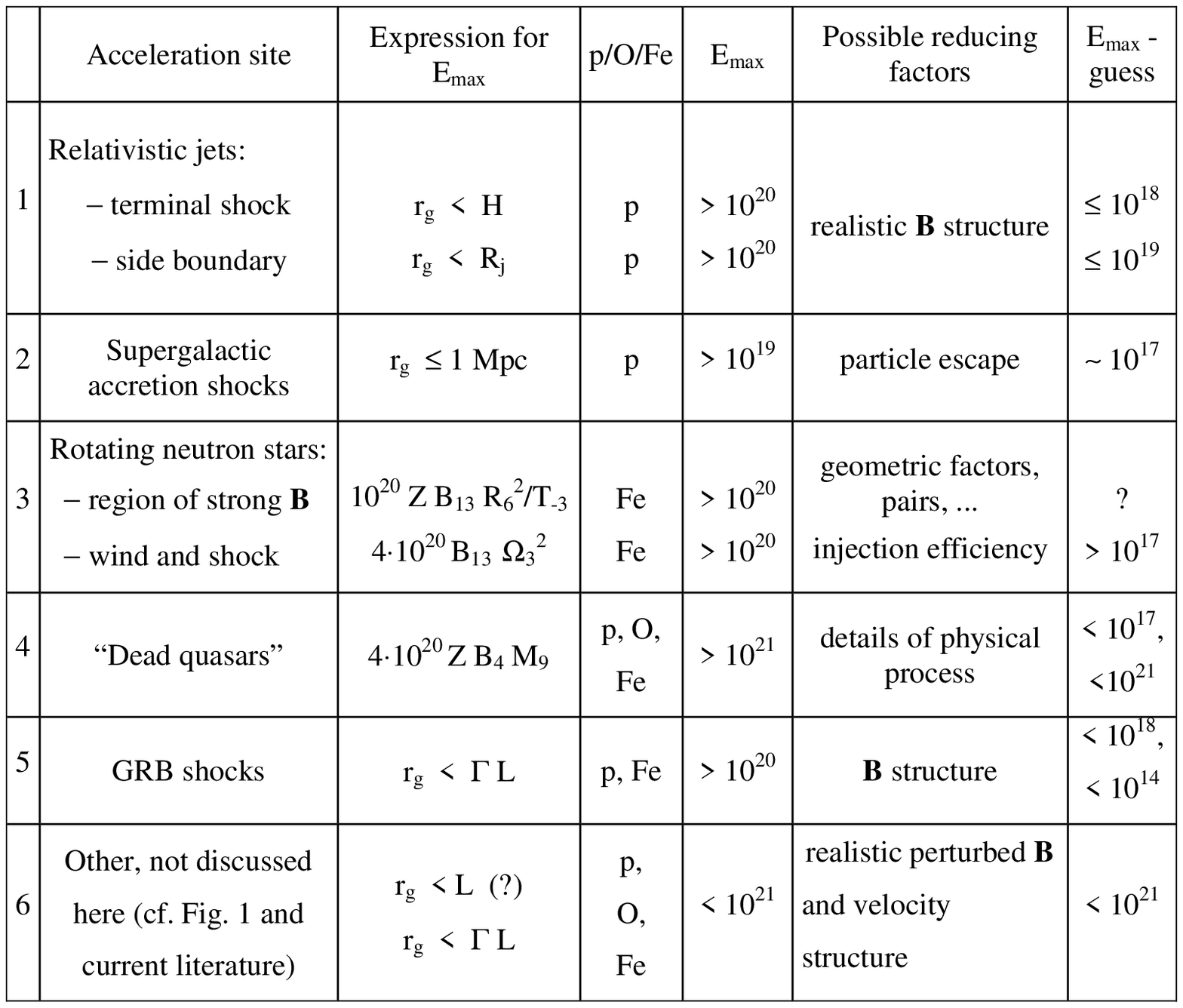}
\caption{A list of UHE CR acceleration sites discussed in the text.
In the third column we list main, order of magnitude conditions for the 
maximum energy of accelerated particles, $E_{max}$. In the next column
 the considered nuclei are mentioned;
here O stands for all medium atomic number nuclei. In the 
last column the author's estimates of $E_{max}$ are provided, taking 
into account his subjective evaluation of various reducing factors 
limiting the acceleration process.}
\end{table}

Ability to accelerate particles up to ultra high energies 
$\sim 10^{20}$~eV by sources (models) discussed shortly in this 
paper (cf. Table 1) and a few further ones appearing in the literature 
is an open 
problem. Very few of these propositions reached even a semi-quantitative 
level, involving a detailed modelling of individual particle 
acceleration. Quite often for the evaluated process one assumes the 
physical conditions providing the most efficient acceleration, these 
conditions are usually only weakly constrained by observations. Besides 
these reservations one should note a growing number of `bottom-up' 
models, using a classical physics to explain UHE CR origin. Possibly one
of these or a new proposed mechanism will prove to work. However, the 
main present day task for UHE CR astrophysics is to increase the number
and quality of observational data, to allow for a more serious
critical evaluation of the considered acceleration models and to discuss
transport processes and forming the observed particle spectrum. In
particular, besides the traditional questions about UHE cosmic rays
(spectrum, anisotropy, what particles we observe), a number of
astrophysical issues must be resolved like understanding of the
intergalactic magnetic field structure, confirmation or not of the
ultrarelativistic winds from pulsars, existence or not of the galactic
winds, direct detection and/or explaining composition of the cosmic dark
matter, etc. When considering chances for understanding UHE cosmic rays
one have to take into account the fact that we are still far from
explaining mechanisms providing much better studied lower energy cosmic
rays.

\section*{Acknowledgements} 
 
I am grateful to Gra\.zyna Siemieniec-Ozi\c{e}b{\l}o, Marek Sikora, 
W{\l}odek Bednarek and Heniek Wilczy\'nski for discussions on UHE CRs, 
for Maria Giller for critical remarks on the manuscript, and for Murat 
Boratav for providing the Hillas plot. Remarks of the 
anonymous referee helped to improve the final version of the paper.
The work was supported by the {\it Komitet Bada\'n Naukowych} within the 
project 2~P03B~112~17 and the grant PB 258/P03/99/17~. 
 
\section*{Appendix}

\begin{figure}
\vspace{20,3cm}
\includegraphics{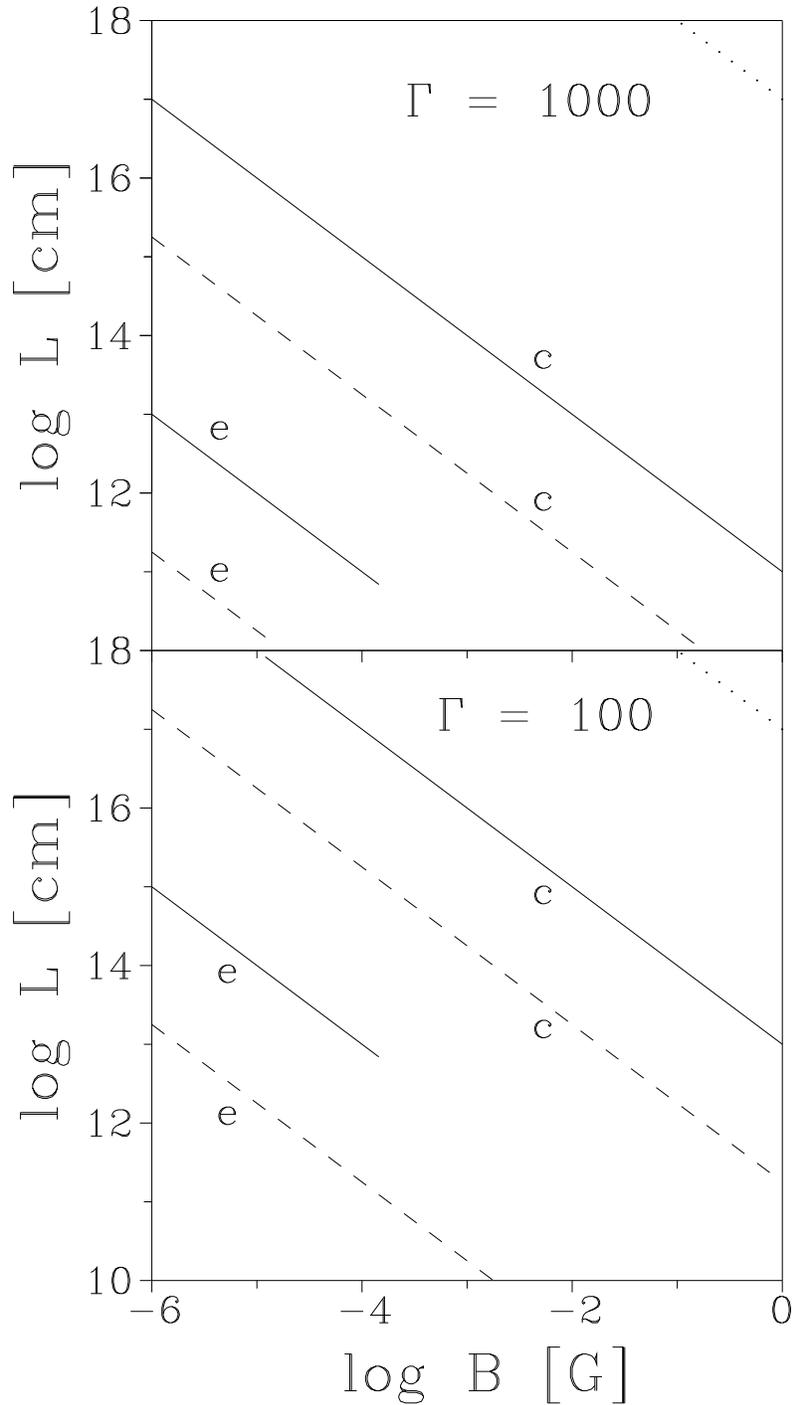}
\caption{A version of the Hillas plot illustrating the possibility of
particle acceleration up to $10^{20}$ eV by reflection from the
ultra-relativistic shock with the Lorentz factor $\Gamma = 10^3$ or
$10^2$. At the axes we present the magnetic field, $B$, upstream of the
shock (as measured in the upstream plasma rest frame) and the shock
size, $L$, as measured along the shock surface, perpendicular to the
flow. $L$ is taken equal to the gyroradius of the highest energy shock
reflected particle, as measured downstream of the shock.}
\end{figure}

To draw a version of the Hillas plot presented on Fig.~2
we consider the situation with the
magnetic field energy density being in equipartition with the plasma
energy density upstream of the shock: the sound velocity in the upstream
plasma is assumed $v_s = 30$ km/s and the respective plasma density is
defined by the equipartition condition. One should note that the
magnetic field downstream of the shock, $B^\prime$, is much stronger
than the upstream value given at the horizontal axis. The compressed
(`c') upstream field reaches a downstream value $\sim 3 \Gamma$ higher,
while in the often considered case of the equipartition (`e') between
the magnetic field and the plasma downstream of the shock the required
field is higher by an additional large factor $\sim c/v_s$. The curves
presented at Fig.~2 -- 
solid lines for protons and dashed lines for Iron nuclei -- are plotted for
particles reaching $10^{20}$ eV after reflection from the shock, as
measured upstream of the shock. To draw the figure each value of the
upstream magnetic field given at the x-axis was transformed downstream
the shock according to the `c' or `e' prescription and a gyroradius
$r_g(B^\prime)$ (Eq. 1) was derived for a particle with the energy
$10^{20} {\rm eV} \, / \, \Gamma$ and the transformed magnetic field
$B^\prime$. This value $r_g(B^\prime) = L$ is given at the y-axis. One
should note that the size $L$ has to be a distance between causally
connected regions along the shock. For example for the spherical shock
with a radius $R$ the condition $L < R/\Gamma$ has to be satisfied. On
the figure, we do not present results for protons with the radiative
loss time scale (Eq.~3) downstream of the shock shorter than the derived
$L/c$. For a reference at the upper right corner of each panel a dotted
line is added representing the $\beta = 1$ line for $10^{20}$ eV protons
at the original Hillas plot: $r_g(B) = L$.

\pagebreak

\end{document}